\documentclass[aps, prab, reprint, groundscriptaddress]{revtex4-2}

\usepackage{amsmath, newtxtext, newtxmath} 
\usepackage{graphicx} 
\usepackage{hyperref}
\hypersetup{
    colorlinks = true,        
    linkcolor = blue,          
    citecolor = blue,        
    filecolor = magenta,      
    urlcolor = blue           
}
\usepackage{color}

\usepackage{siunitx} 

\begin{document}

\title{Generating 77 T using a portable pulse magnet for single-shot quantum beam experiments}

\author{Akihiko~Ikeda}
\email[]{a-ikeda@uec.ac.jp}
\affiliation{Department of Engineering Science, University of Electro-Communications, Chofu, Tokyo 182-8585, Japan}
\author{Yasuhiro~H.~Matsuda}
\author{Xuguang~Zhou}
\author{Shiyue~Peng}
\author{Yuto~Ishii}
\author{Takeshi~Yajima}
\affiliation{Institute for Solid State Physics, University of Tokyo, Kashiwa, Chiba 277-8581, Japan}
\author{Yuya~Kubota}
\author{Ichiro~Inoue}
\affiliation{RIKEN SPring-8 Center, Kouto, Soyo, Hyogo 679-5198, Japan}
\author{Yuichi~Inubushi}
\author{Kensuke~Tono}
\author{Makina~Yabashi}
\affiliation{RIKEN SPring-8 Center, Kouto, Soyo, Hyogo 679-5198, Japan}
\affiliation{Japan Synchrotron Radiation Research Institute, Kouto, Soyo, Hyogo 679-5198, Japan}

\date{\today}

\begin{abstract}
We devised a portable system that generates pulsed high magnetic fields up to 77 T with 3 $\mu$s duration.
The system employs the single turn coil method, a destructive way of field generation.
The system consists of a capacitor of 10.4 $\mu$F, a 30 kV charger, a mono air-gap switch, a triggering system, and a magnet clamp, which weighs less than 1.0 tons in total and is transportable.
The system offers opportunities for single-shot experiments at ultrahigh magnetic fields in combinations with novel quantum beams.
The single-shot x-ray diffraction experiment using x-ray free-electron laser at 65 T is presented.
We comment on the possible update of the system for the generation of 100 T.
\end{abstract}

\maketitle

\begin{figure*}
\includegraphics[width = \linewidth, clip]{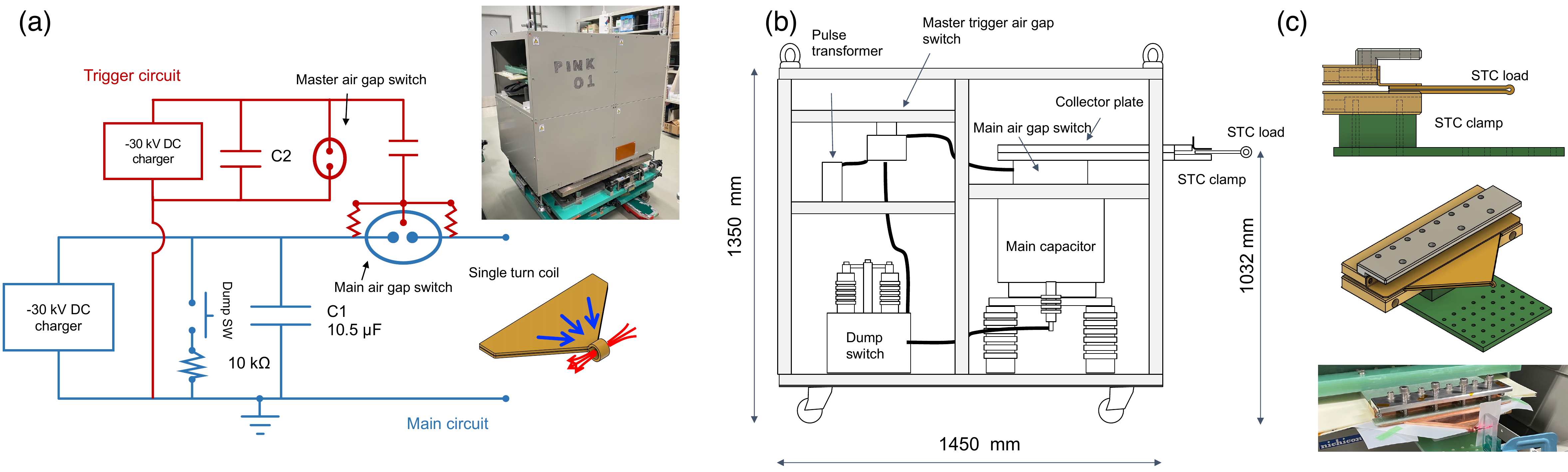} 
\caption{
(a) Schematic drawing of the electric circuit of PINK-01, where the main circuit and the trigger circuit are shown.
(b) Schematic drawing of the side view of PINK-01, where the arrangement of the parts is shown.
(c) Drawing of the magnet clamp, STC, and stage shown with a photo.
\label{setup}}
\end{figure*}

\begin{figure*}
\includegraphics[width =  \linewidth, clip]{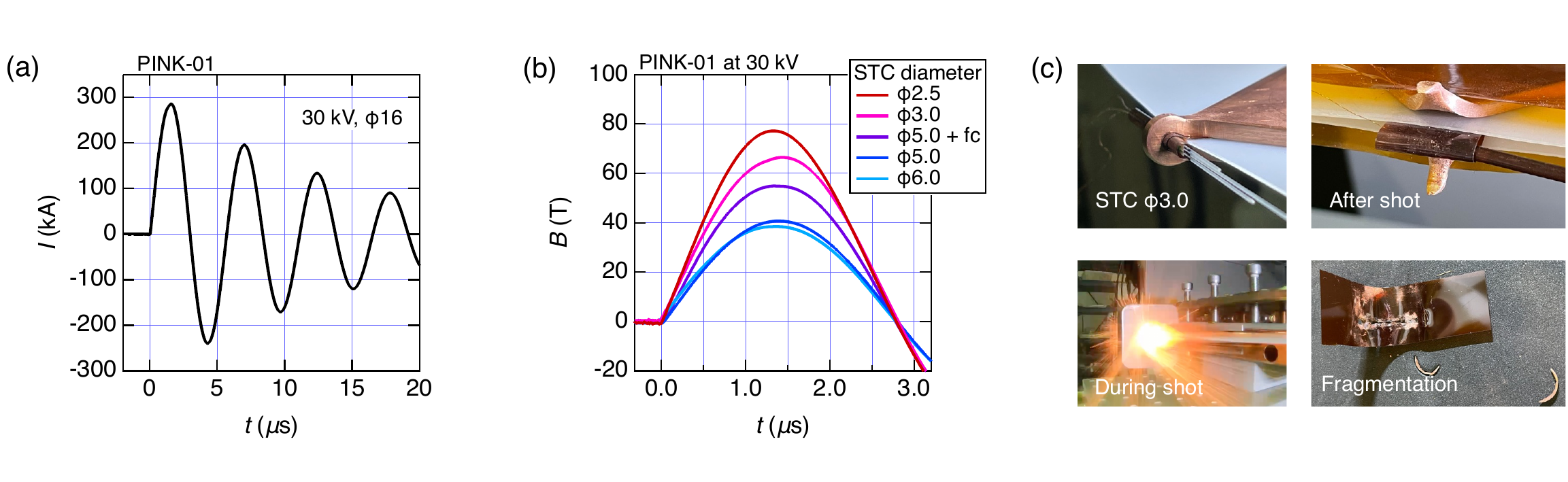} 
\caption{
(a) The current waveform of PINK-01 discharged with a charging voltage of 30 kV where the load is an STC of $\phi$16 mm.
(b) The magnetic field waveform generated using PINK-01 discharged with a charging voltage of 30 kV where the STC loads of $\phi$2.5 to 6.0 mm are used. FC stands for the flux concentrator.
(c) Photos of the STC of $\phi$3.0 mm before, after, and during the pulsed magnetic field generation up to 65 T. The bottom right photo shows the fragmentation of the coil produced in the pulsed-field generation.
\label{discharge}}
\end{figure*}

\begin{figure*}
\includegraphics[width =  \linewidth, clip]{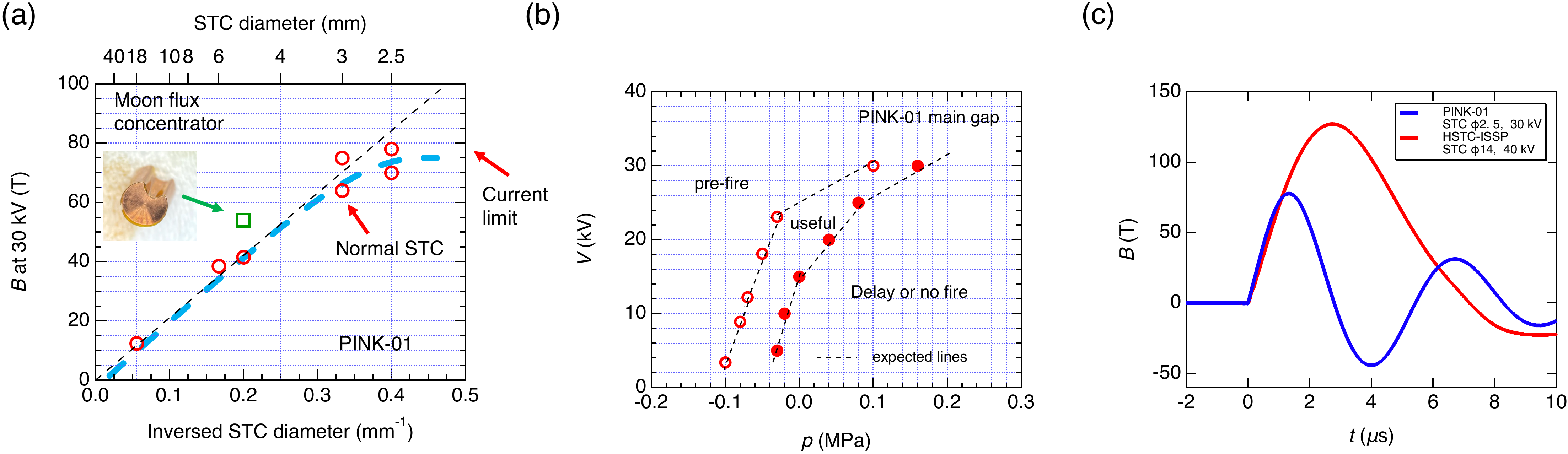} 
\caption{
(a) The dependence of the maximum field generated using PINK-01 discharged with a charging voltage of 30 kV on the STC diameter. The inset shows a flux concentrator with a shape of a crescent Moon being used with an STC of $\phi$5 mm. A green square represents the data for the flux concentrator.
(b) The dry air pressure dependence of the characteristics of the main air-gap switch investigated for the range of charging voltages from 3 to 30 kV.
The air pressure range between filled and open circles is useful for a successful discharge.
The higher pressure or lower pressure result in the delay, no fire, or pre-firing of the discharge, respectively. 
(c) The magnetic field waveforms of PINK-01 and the horizontal STC at ISSP, Tokyo, for comparison.
\label{stcother}}
\end{figure*}

\begin{figure}
\includegraphics[width = 0.8\linewidth, clip]{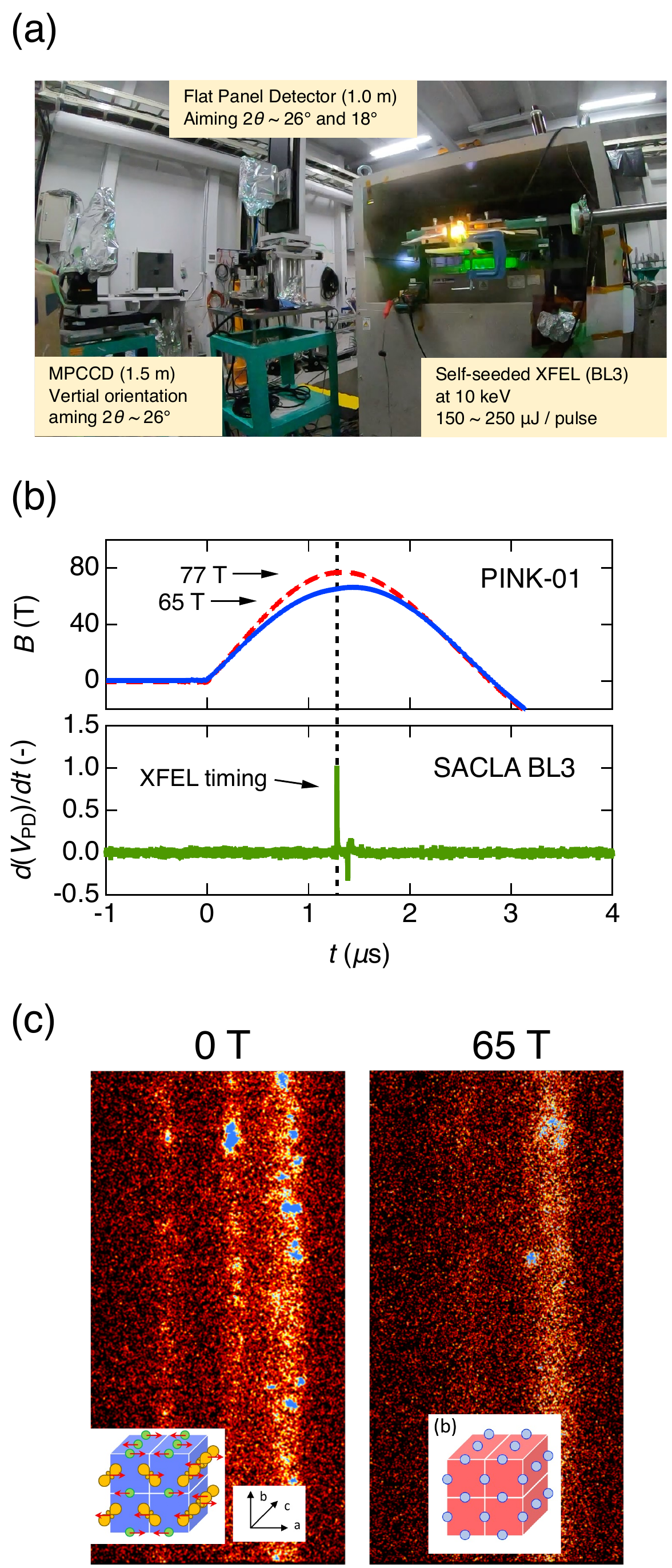} 
\caption{
(a) A photo of the experimental set up of the pulsed high magnetic field experiment using PINK-01 and SACLA.
(b) The magnetic field waveform of PINK-01 with the STC of $\phi$2.5 mm at a charging voltage of 30 kV, which is synchronized with the XFEL pulse at BL3 of SACLA, Hyogo, Japan.
The bottom figure shows the XFEL timing detected using a photodiode.
(c) A part of the image of the Debye Scherrer ring detected using MPCCD in SACLA (Left) before shot and (Right) during the magnetic field pulse of 65 T, where the field-induced lattice change is clearly shown.
The sample is Bi$_{0.5}$Ca$_{0.5}$MnO$_{3}$ at room temperature.
\label{sacla}}
\end{figure}

Magnetic field is an indispensable thermodynamic parameter in condensed matter physics, that acts directly on the magnetic moment of the spin and orbital angular momentum of the electrons.
The strength and the direction of the magnetic field are useful control parameters. 
With a high magnetic field, one can induce kinds of phase transitions in materials.
This is reasonable considering the Zeeman energy splitting  of an electron spin at 100 T amounts to 134 K ($ \simeq 11.5$ meV).
One spectacular example is the finding of the novel $\theta$ phase of solid oxygen, where the antiferromagnetic ordering of the molecular spins are transformed into a ferromagnetic one, resulting in the collapse of the molecular arrangement and crystal structure, leading to the appearance of the new phase \cite{NomuraPRL2014}.
One finds recent discovery at above 100 T not only in the strongly correlated electron systems $\rm{YbB_{12}}$ \cite{TerashimaJPSJ2017}, FeSi \cite{NakamuraPRL2021}, $\rm{VO_{2}}$ \cite{MatsudaNC2020}, $\rm{LaCoO_{3}}$ \cite{IkedaPRB2016, IkedaPRL2020, IkedaArXiv} but also in the quantum spin systems \cite{MatsudaPRL2013}, frustrated magnets pyrochlore systems \cite{MiyataPRL2011} and breathing pyrochlore systems \cite{GenArXiv}.

 For the magnetic field generation well above 100 T, one needs a destructive pulse magnet, because the electromagnetic force well above the destruction strength of the magnet wire is applied when generating over 100 T \cite{HerlachRPP1999}.
 Destructive pulse magnets, such as the single turn coil (STC) method and the electromagnetic flux compression (EMFC) methods \cite{NakamuraRSI2018} are used to generate 100 - 300 T and 300 - 1200 T, respectively.
 The characteristic of the destructive pulse magnets is that typically the pulse duration of the destructive pulse magnets is as short as a few micron seconds.
Despite the harsh experimental condition in the destructive pulse magnets such as short pulse duration, electric noises, and single-shot experiment, many experimental techniques have been successfully developed, that range from magnetoresistance \cite{SuyeonJPSJ2015, KohamaRSI2020, NakamuraMST2018}, magnetostriction \cite{IkedaRSI2017}, to ultrasound measurements \cite{NomuraRSI2021}.

Macroscopic and microscopic probes complement each other in condensed matter studies are thus indispensable.
 However, microscopic probes of the lattice, electronic states, and magnetic states, such as neutron scattering, x-ray diffraction, spectroscopy, and resonance techniques are elusive with pulsed high magnetic fields because of the shortage of the accumulation time of the signal. 
 Below 100 T generated with millisecond pulses, there have been great efforts such as x-ray experiments \cite{MatsudaJPSJ2013}, nuclear magnetic resonance \cite{IharaRSI2021}, teraheltz time-domain spectroscopy (THz-TDS) \cite{NoeRSI2013}.
However, these microscopic measurements are impossible at above 100 T because 100 T environments are limited to the pulse durations of a few $\mu$-seconds. 
 
 The situation has been altered by the emergence of the highly bright and short-pulsed x-ray source, x-ray free-electron laser (XFEL) has enabled single-shot x-ray experiment in a few tens of femtoseconds \cite{BostedtRMP2016, SeddonRPP2017}.
 Such a short-pulsed single-shot experiment is compatible with the  $\mu$-second pulsed ultrahigh magnetic fields.
Phase transitions under high magnetic fields involve a variety of structural changes harvesting a wealth of physics \cite{IkedaPRR2020}, via the valence change, spin-state change, electron-lattice couplings in metal-insulator transitions, spin-lattice couplings in magnetic order changes.
Thus, a small portable high magnetic field generator is in demand.
Previous portable pulse magnets have been limited below 50 T.
At present, 100 T generators are always very big and not portable.

Here we developed a portable pulsed power system for high magnetic field generation.
The system is called PINK-01 ({\bf P}ortable {\bf IN}tense {\bf K}yokugenjiba) which is a destructive pulse magnet using the STC method.
PINK-01 is portable and be able to generate a pulsed high magnetic field of 77 T.
 We describe the performance of the magnetic field generation.
 We also discuss the application of PINK-01 to single-shot quantum beam experiments such as XFEL.
 We further discuss the possible future applications and updates.

PINK-01 consists of the main capacitor of 10.4 $\mu$F, whose rated voltage is 30 kV storing the total energy of 4.5 kJ.
To allow the fast discharge, we employed an air-gap switch (Main air-gap), which is triggered by another air-gap switch (Master air-gap).
The main circuit contains no cable, where the air-gap switch is directly housed on top of the capacitor and the collector plate guides the current to the STC load.
The system possesses the dump switch for safety.
The circuit of PINK-01 is schematically shown in Fig. \ref{setup}(a).
The dimension of the Faraday cage of PINK-01 is $1350 \times 1450 \times 1000$ mm, whose side view is shown in Fig. \ref{setup}(b).
The total weight of the main box of PINK-01 is 650 kg.
PINK-01 is equipped with casters and shackles, which makes it transportable.
The magnet clamp is designed so that the electric contact to the load STC is ensured as shown in Fig. \ref{setup}(c).
The whole setup of the pulse power system of PINK-01 excluding the magnet clamp and the load STC is manufactured by NICHICON.

A typical waveform of the discharged current is shown in Fig. \ref{discharge}(a) where the maximum current of 280 kA is obtained with the pulse duration of 3.0 $\mu$s.
The STC of $\phi$16 mm is used as a load.
The current waveform is well fitted with the analytical form of the damped oscillation with a fitting parameters of $R = 13$ \si{\mohm}, $L =$ 80 nH, 10.4 $\mu$F, 30 kV.
Considering $L_{\rm{STC}}\simeq 10$ nH, the residual inductance is $L_{\rm{res}} \simeq$ 70 nH.
The waveforms of the magnetic field generated using the STC loads of $\phi$6.0 mm to 2.5 mm are shown in Fig. \ref{discharge}(b), where the maximum fields increase with decreasing diameter of the STCs.
The maximum field of 77 T is obtained using the STC of $\phi$2.5 mm.
The generation of the field 40 T accompanies the deformation of the STC.
The generation over 60 T accompanies the explosion of the STCs.
The photo of the STC $\phi$3.0 mm before, after and during the shot are shown in Figs. \ref{discharge}(a) - \ref{discharge}(d).

The maximum magnetic fields are investigated as a function of the inner diameter of the STC loads as shown in Fig. \ref{stcother}(a).
The maximum magnetic field linearly increases with the inverse diameter of the STCs, which is in good agreement with the expectation from Ampere's law.
However, the linearity is lost and the possible maximum field is reached at $\phi$2.5 mm.
This might happen because the allowable maximum current density is reached at $\phi$2.5 mm, where the Joule heating at the STC head is so rapid that the evaporative deformation of the coil happens earlier than that with larger coils.

Jitter is less than 0.1 $\mu$s, which ensures good compatibility with the femtosecond single-shot pulses with the field variation less than 1 $\%$.
The discharge behavior of PINK-01 is strongly dependent on the condition of the air-gap switches.
In the present system, there is a sufficiently large range of the optimum air pressure both for the triggering master air-gap switch and the main air-gap switch.
When the air pressure is too low or too high, it results in the pre-fire or no-fire behaviors, respectively, as shown in Fig. \ref{stcother}(b).
When the air pressure is optimum, we obtain the favorable jitter of less than 0.1 $\mu$s.

For comparison, we show the waveform of the magnetic field generated using the horizontal STC system in the Institute for Solid State Physics in Japan and a STC of $\phi$14 mm as shown in Fig. \ref{stcother}(c).
PINK-01 produces the pulse duration half of that of the HSTC in ISSP.
The ISSP bank is 160 $\mu$F rated at 50 kV storing the energy of 200 kJ.
Roughly, PINK-01 is 1/100 times smaller but the peak field is only 1/10.

Lastly, we discuss the application of PINK-01 for the single-shot quantum beam experiments.
Fig. \ref{sacla}(a) shows a photo of the experimental set up.
Fig. \ref{sacla}(b) shows the magnetic field pulse and synchronized XFEL timing detected at SACLA BL3, the XFEL facility in Japan \cite{IshikawaNP}.
The seeded beam at $h\nu = 10 $ keV has been employed \cite{InoueNP}.
Fig. \ref{sacla}(c) shows a part of Debye-Scherrer rings from a manganite $\rm{Bi_{0.5}Ca_{0.5}MnO_{3}}$ measured before the field generation and on top of the magnetic field pulse of 65 T, respectively, being detected using a multiport charge-coupled device (MPCCD) \cite{Kameshima} placed at $2\theta \simeq 26^{\circ}$.
One can clearly see the change of the XRD pattern from that at 0 T to that at 65 T, where three peaks changes to 2 peaks.
The series of the experimental results and the quantitative analysis are beyond the scope of the present paper and will be reported elsewhere.

Presently the maximum fields of 77 T - 65 T are obtained with STC of $\phi$2.5 mm - 3.0 mm.
As discussed earlier, one can not use an even smaller coil for 100 T generation, because of the possible heating of the coil head.
For higher magnetic field generation, a higher current than the present 300 kA is needed.
Besides, for the condensed matter experiment at low temperatures, a larger bore size of at least $\phi$5 mm is favorable.
To maintain the maximum magnetic field with a larger magnet bore, one needs a higher current, too.
To produce a higher magnetic field with larger STCs, one strategy is to increase the number of capacitor banks and air-gap switches.
In increasing the number of capacitors, possible technical difficulties lie in the synchronization of the air-gap switches and also the increasing size of the system.
Thus, we need to make the system separatable to keep the size and weight of each component transportable.
These are the future direction to a new PINK-02.

Previous STC systems were built in Illinois \cite{HerlachJPE1973, HerlachIEEE1971}, in Tokyo \cite{NakaoJPE1985}, in Chiba \cite{MiuraJLTP2003}, in Berlin \cite{PortugallJPD1997, PortugallJPD1999}.
Among them, two STC systems in Chiba, one STC in Berlin (now in Toulouse, France) and one STC system in Los Alamos systems are in operation, which are all large systems generating over 200 T.
A portable STC system is reported in Ref. \cite{PortugallJPD1997}, where two capacitors of 6 $\mu$F rated at 60 kV and two rail gap switches rated at 750 kA each are used to generate 200 T using an STC of $\phi$5 mm.
Another STC system is reported Ref. \cite{HerlachIEEE1971} using 20 capacitors of 280 $\mu$F in total are assembled in Stanford Linear Accelerator Center, where 150 T using STC method and 100 T using EMFC method are generated in synchronization with the high energy electron beam.
In comparison to those systems, PINK-01 is characterized by a low voltage of 30 kV and a small current of 0.3 MA.
We believe that the successful application to the XFEL experiment in the present study is owing to the moderate scale of PINK-01, which made it compatible with the sophisticated electronics in the experimental hatch.

As another promising application, we finally comment on the possible TDS-THz experiment using PINK-01.
Recently, single-shot TDS-THz experiments are reported where TDS-THz spectroscopy can be carried out in a single pulse of a few femtoseconds \cite{MinamiAPL2013, TeoRSI2015, NoeOE2016}.
The applicability and portability of PINK-01 may make it possible to conduct TDS-THz at 100 T.

 A portable high magnetic field generator  PINK-01 has been developed.
A pulsed high magnetic field of 77 T with 3 $\mu$s time duration has been generated.
The field is compatible with an XFEL for an XRD study of the magnetic field-induced phase transition up to 77 T.
PINK-01 opens a novel venue for single-shot quantum beam experiments at the 100 T range.

\begin{acknowledgements}
This work was supported by the Basic Science Program Grant No. 18-001 of TEPCO Memorial Foundation and JSPS KAKENHI Challenging Research (Pioneering) No. 20K20521.
The present experiment was performed with the approval of the
Japan Synchrotron Radiation Research Institute (Proposal No.
2021A8063).
The authors would like to acknowledge the supporting members of the SACLA facility.
\end{acknowledgements}

\bibliography{pink01}
\end{document}